\documentstyle[11pt,aaspp4]{article}

\font\caps=cmcsc10 scaled 1200
\def\etal {et~al.}
\def\radm2{radians m$^{-2}$}
\def\deg   {{$^\circ$}}

\def\HI {H\kern0.1em{\sc i}}

\slugcomment{Accepted to the Astrophysical Journal for v. 519 July 1999}

\lefthead{Taylor, Silver, Ulvestad, \& Carilli}
\righthead{Star Formation in Mrk\,231 }

\begin{document}

\title{~~\\ ~~\\ The Starburst in the Central Kiloparsec of Markarian 231 }

\author{G.~B.~Taylor, C.~S.~Silver, J.~S.~Ulvestad, \& C.~L.~Carilli;
gtaylor,csilver,julvestad,ccarilli@nrao.edu}
\affil{National Radio Astronomy Observatory, \\
       P.O. Box O, Socorro, NM 87801, USA}
\authoremail{gtaylor@nrao.edu, csilver@nrao.edu, julvesta@nrao.edu,
  ccarilli@nrao.edu}

\begin{abstract}

  We present VLBA observations at 0.33 and 0.61 GHz, and VLA
  observations between 5 and 22 GHz, of subkiloparsec scale radio
  emission from Mrk\,231.  In addition to jet components clearly
  associated with the AGN, we also find a smooth extended component of
  size $100 - 1000$ pc most probably related to the purported massive
  star forming disk in Mrk\,231.  The diffuse radio emission from the
  disk is found to have a steep spectrum at high frequencies,
  characteristic of optically thin synchrotron emission.  The required
  relativistic particle density in the disk can be produced by a star
  formation rate of 220 M$_\odot$ yr$^{-1}$ in the central kiloparsec.
  At low frequencies the disk is absorbed, most likely by ionized gas
  with an emission measure of $8 \times 10^5$ pc cm$^{-6}$.  We have
  also identified 4 candidate radio supernovae that, if confirmed,
  represent direct evidence for ongoing star formation in the central
  kiloparsec.

\end{abstract}

\keywords{galaxies:active --- galaxies:individual(Mrk\,231) --- galaxies: jets --- galaxies:ISM --- galaxies: nuclei --- galaxies:halos}


\section{Introduction}

Mrk\,231 is one of the most luminous infrared galaxies in the local (z
$<$ 0.1) universe (Surace et al.\ 1998).  As such it is potentially a
key object for understanding the mechanism(s) that power such extreme
far infrared dust emission.  The most likely dust heating mechanisms
are either (1) radiation from an AGN; or (2) a powerful starburst.
The existence of an AGN in Mrk\,231 is clearly demonstrated by the
presence of a parsec-scale jet (Neff \& Ulvestad 1988, Ulvestad,
Wrobel, \& Carilli 1999a), but the extent to which the AGN powers the
high infrared luminosity (10$^{12}$L$_\odot$) is unclear.  Based on CO
observations, Downes \& Solomon (1998) suggest that most of the FIR
luminosity in the central kpc of Mrk\,231 comes from a starburst.
Other evidence for a starburst is the large amount of dust and
extinction (Cutri, Rieke, \& Lebofsky 1984).  Based on its
asymmetrical optical morphology with tidal tails, a merger in the last
10$^9$ years is likely (Hamilton \& Keel 1987; Hutchings \& Neff 1987;
Armus \etal\ 1994; Surace \etal\ 1998).  Even if the merger is not a
direct energy source for the high infrared luminosity, it may play a
critical role in channeling large amounts of gas and dust into the
central kiloparsec, which then either feed the AGN or trigger the
starburst.

Carilli, Wrobel, \& Ulvestad (1998) report the discovery of a 
subkiloparsec scale disk seen in both the centimeter radio continuum
and in neutral atomic hydrogen.  Such a structure is almost 
unknown among AGN.  Only one other radio source, 3C\,84, has a
$\sim$100 pc disk or millihalo surrounding the parsec-scale jets
(Silver, Taylor, \& Vermeulen 1998).  In the case of 3C\,84, 
the most likely origin for the millihalo is from relativistic
particles that have diffused out from the parsec-scale jets, but
a starburst origin cannot be completely ruled out.  Carilli \etal\
suggested that the millihalo seen around 
Mrk\,231 is really a disk with a massive star formation rate of
order 100 M$_\odot$ yr$^{-1}$.  Here we present further VLBA 
observations of the disk in Mrk\,231 at 0.61 and 0.33 GHz in an
attempt to learn more about its spectrum and structure.  We have
also reduced and analyzed archival VLA observations which reveal
the outer part of the disk at 5 -- 22 GHz.

We assume H$_0$ = 50 km s$^{-1}$ Mpc$^{-1}$ and q$_0$ = 0.5
throughout, so at the redshift of Mrk\,231 of 0.04152 (de Vaucouleurs
\etal\ 1991), 1\arcsec\ corresponds to 1.12 kpc.

\section{VLBA Observations and Data Reduction}

The VLBA observations were carried out at 0.333 and 0.613 GHz with the
10-element VLBA of the NRAO\footnote{The National Radio Astronomy
Observatory is operated by Associated Universities, Inc., under
cooperative agreement with the National Science Foundation} in a
single 10-hour observing session on 1998 June 20. The net integration
time on Mrk\,231 was 250 minutes at each frequency spread over
multiple snapshots to improve $u, v$ coverage. Both frequencies were
observed simultaneously with 3 IF bands covering the frequency range
0.328--0.340 GHz and 1 IF covering the range 0.611--0.615 GHz.  Both right- and
left-circular polarizations were observed with 2-bit sampling. Due to
strong external interference near 0.613 GHz a 3.4 MHz passband filter was
used on all antennas. The correlator produced a 2.06 second
integration time and 16 channels per IF band.

Standard {\it a priori} flagging, amplitude calibration, fringe fitting,
bandpass calibration, and frequency averaging procedures were followed
in AIPS.  Phase referencing was performed by recorrelating at the
position of the quasar  J1252+5634 located some 36\arcmin\ from 
Mrk\,231.  Global fringe fitting was performed on J1252+5634 and
the resulting delays, rates and phases transfered to Mrk\,231. 
Subsequently, Mrk\,231 was phase-only self-calibrated with a 2
minute solution interval.
All manual editing, imaging, deconvolution, and
self-calibration were performed using {\caps Difmap} (Shepherd,
Pearson, \& Taylor 1994; Shepherd 1997). 

In Fig.~1 we show the 0.33 GHz and 0.61 GHz VLBA images at a
resolution of 30 mas.  Both images are dominated by a compact source,
just resolved in the north-south direction.  Due primarily to the
greater bandwidth, the 0.33 GHz image is more sensitive by a factor of
$\sim$2.

We have also reanalyzed the 1.36 GHz continuum data of Carilli \etal\ 
(1998).  By performing careful editing and self-calibration within
{\caps Difmap} we were able to achieve an image with a rms noise of 0.02
mJy/beam (Fig. 2).  The Pie Town -- VLA baseline was removed, eliminating
the extended inner disk emissions which are not easily deconvolved 
and otherwise severely limit the achievable dynamic range.

\begin{figure}
\vspace{16cm}
\includegraphics{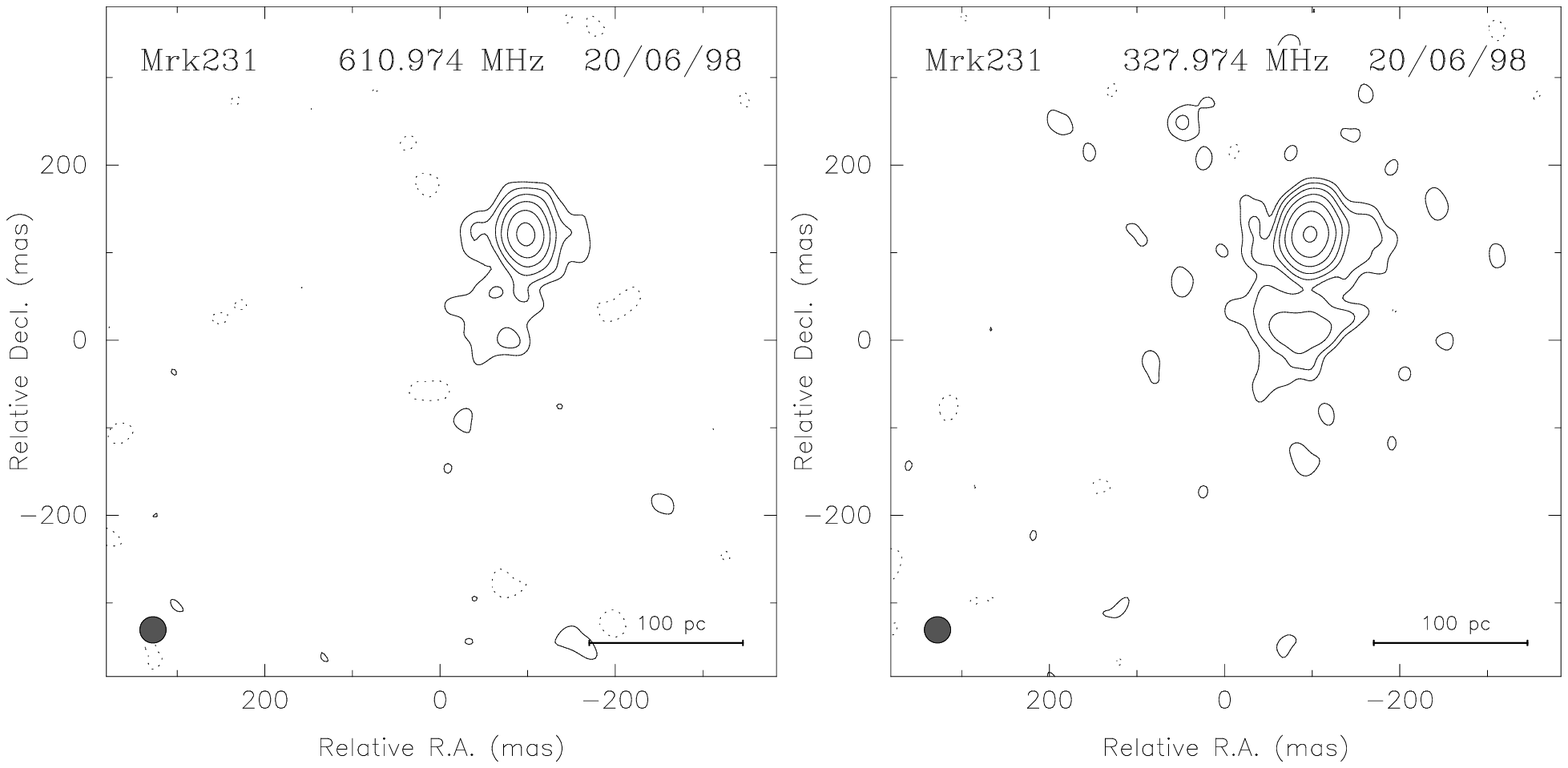}
\figcaption{The naturally weighted and tapered images of Mrk\,231 at 0.61
and 0.33 GHz. The resolution of both images is 30 mas.  Contours are at 
factor 2 intervals and start at 2 and 1 mJy/beam for 
0.61 and 0.33 GHz respectively.
\label{fig1}}
\end{figure}

\placefigure{fig1}

\begin{figure}
\vspace{16cm}
\includegraphics{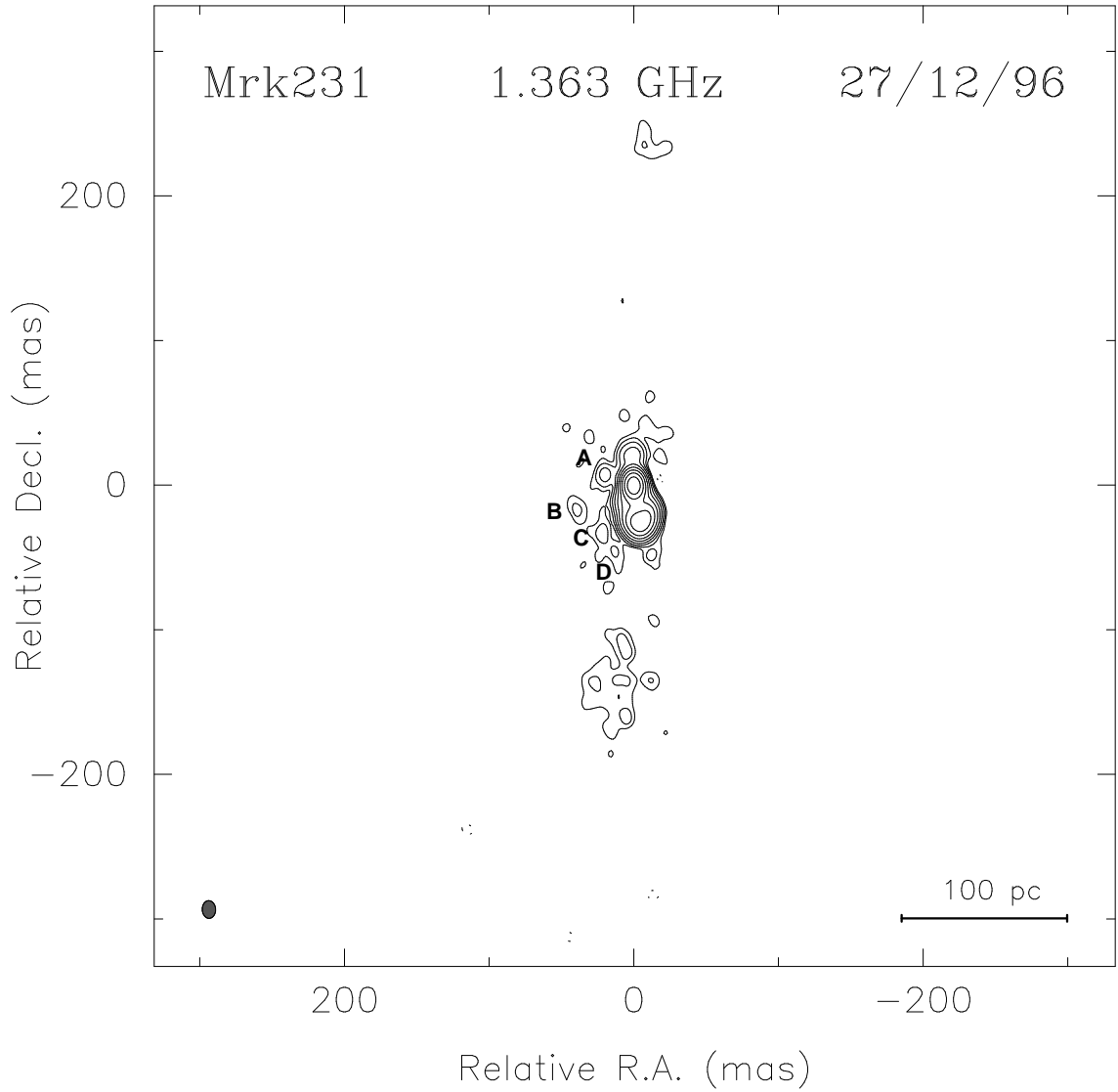}
\figcaption{The 1.3 GHz continuum data of Carilli \etal\ (1998) reimaged
with a naturally weighted beam of 12.3 $\times$ 9.3 mas in
p.a.\ 3$^\circ$. 
Contours are drawn at 
factor 2 intervals and start at 0.1 mJy/beam (5$\sigma$).  Four RSNe candidates
are labeled A--D.
\label{fig2}}
\end{figure}

\placefigure{fig2}



\section{VLA Observations and Data Reduction}

All the VLA observations reported here are A configuration
observations retrieved in raw form from the VLA archive.  These
observations were carried out using the VLA in the standard fashion
with a 100 MHz continuum observing mode at 4.86, 8.44, 14.94 and
22.23 GHz at epoch 1995.55, 1991.48, 1983.87, and 1996.99 respectively. 
The total observing time on Mrk\,231 was 210, 15, 35, and 90 minutes 
at 5--22 GHz respectively.

Once the data were calibrated in AIPS, they were exported to {\caps Difmap}
for further self-calibration.  To better reveal the underlying
extended emission, a delta function was subtracted from the phase
center of each observation.  The resulting naturally weighted images
are shown in Fig.~3.  The amplitude of the delta function subtracted
represents the emission from the VLBI triple seen on scales of
$\sim$40 mas (Ulvestad \etal\ 1999a), and the 30 -- 100 mas radius
inner disk component first detected by Carilli \etal\ (1998).
This sum is labeled the ``nucleus'' in Table 1 and in Fig.~4. The
remaining flux density in the images (radii of 0.1 -- 1\arcsec ), is
labeled the ``outer disk'' in Table 1.

\begin{center}
TABLE 1 \\
\smallskip
C{\sc omponent} F{\sc lux} D{\sc ensities} {\sc and} S{\sc pectra}

\begin{tabular}{l r r r r r r r r r}
\hline
\hline
Component     & $S_{22}$   & $S_{15}$ & $S_{8.4}$   & $S_5$ & $S_{1.3}$ & $S_{0.61}$ & $S_{0.33}$ \\
 & (mJy) &  (mJy) &  (mJy)  & (mJy) &  (mJy) &  (mJy) & (mJy) \\
\hline
\noalign{\vskip2pt}
C+N+S+ID      & 59$\pm$9   & 110$\pm$11  & 208$\pm$15  & 269$\pm$14  & 230$\pm$12 & -- & -- \\
C+N+S         & --   &  62$\pm$6  & 142$\pm$14  & 173$\pm$17   &  99$\pm$10 & 163$\pm$16 & 152$\pm$15 \\
S. Jet        & --   & --   & --   &  --   &  -- &  28$\pm$3 &  44$\pm$4 \\
Inner Disk    & --   &  48$\pm$5  &  66$\pm$7  &  96$\pm$10   & 131$\pm$13 & --  &  30$\pm$6 \\
Outer Disk    & 13.1$\pm$0.7 & 25.3$\pm$1.3 & 32.7$\pm$1.7 &  25.9$\pm$1.3 & -- & -- & --  \\
\hline
\label{tab1}\end{tabular}
\end{center}

\begin{figure}
\vspace{16cm}
\includegraphics{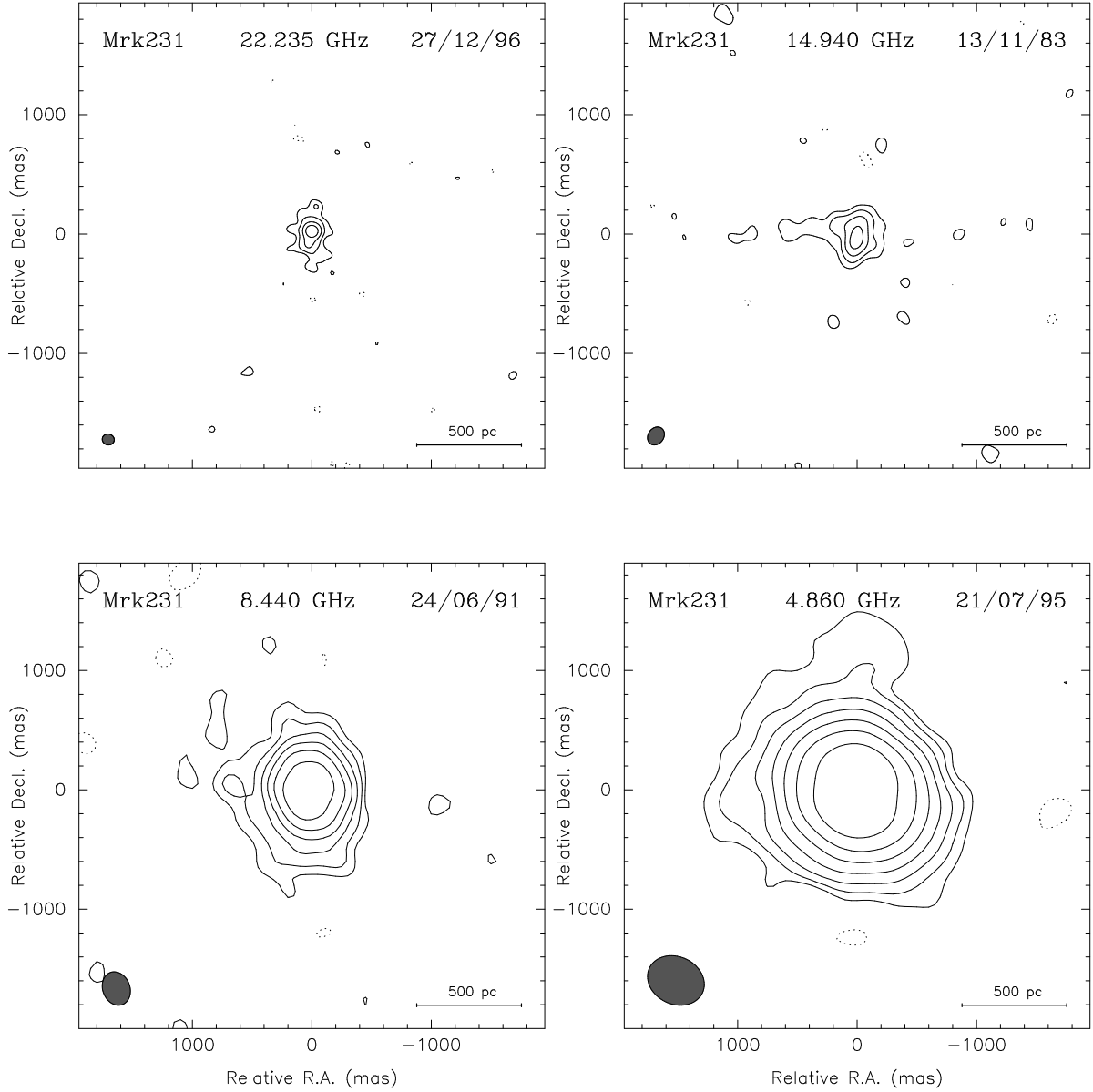}
\figcaption{VLA images of Mrk\,231 at 22, 15, 8 and 5 GHz.  A delta
function has been removed before imaging as described in the text.  
The beam shapes are shown in the lower left corner.  Contours are at 
factor 2 intervals and start at 0.5, 0.8, 0.12 and 0.08 mJy/beam for 
22, 15, 8 and 5 GHz respectively.
\label{fig3}}
\end{figure}

\placefigure{fig3}

\begin{figure}
\vspace{16cm}
\includegraphics{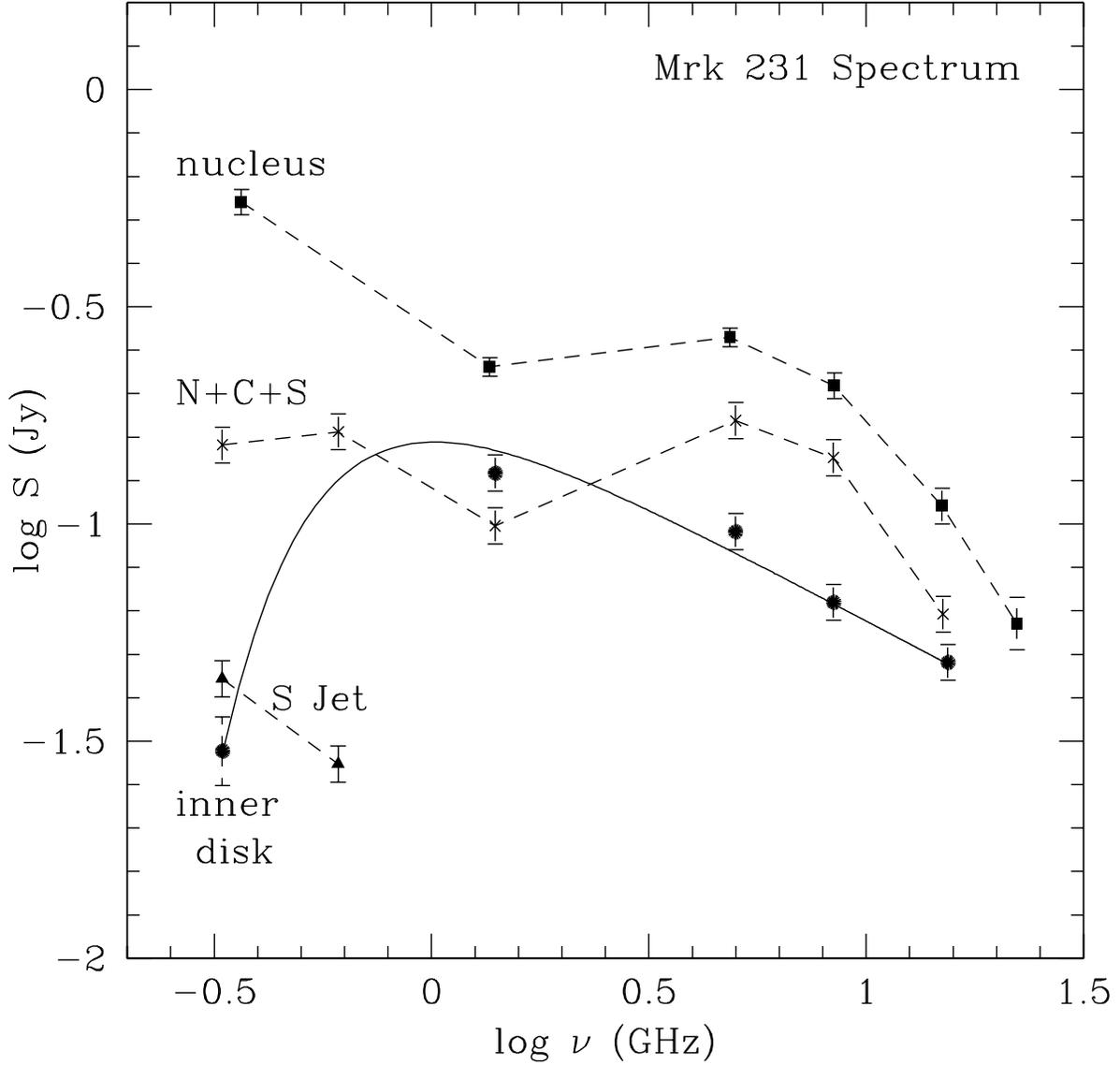}
\figcaption{The spectrum of the constituent components of Mrk\,231 on
  subkiloparsec scales.  The nucleus (squares) refers to the VLA core
  which is presumed to be the sum of the VLBI triple (N+C+S --
  crosses) and the disk (filled circles).  The triangles denote
  measurements of a newly detected southern jet component.
  Measurements of these components come from Ulvestad \etal\ (1999)
  and this paper, except for the flux density of the nucleus at 0.365
  GHz, which is from Douglas \etal\ (1996).  A free-free absorption
  model (solid line) is fit to the disk emission as discussed in the
  text.
\label{fig4}}
\end{figure}

\placefigure{fig4}

\section{The Jet Components}

The dominant ``core'' component in our VLBA images is really comprised
of a north-south triple (Ulvestad \etal\ 1999a).  This triple includes
a GHz peaked spectrum core between two steep spectrum lobes with the
southern lobe being the brighter of the two.  The total extent is
about 60 mas (70 pc).  This structure is just resolved in our 0.61 and
0.33 GHz images.  At 1.3 GHz, the spectra of parts of the N and S
lobes are turning over, while the spectrum of the core turns over near
8 GHz (Ulvestad \etal\ 1999a).  As these are the dominant components,
on that basis one might have expected very little flux density at 0.33
GHz.  We see from our Fig.~4 that the spectrum of the core
conglomerate (N+C+S at this resolution of 30 mas) is essentially flat
out to 0.33 GHz.  This may be due in part to a steep spectrum
contribution from the jet that connects both lobes to the core hinted
at in Fig.~1a of Ulvestad \etal\  The absence of dominant free-free
opacity indicates that at least some parts of the nuclear region have
a relatively unobstructed line-of-sight to the observer, perhaps
cleared out by the AGN.

A new component is seen in our VLBA images some 107 mas south of the
core in position angle 171\deg.  This component is curved into a
bowshock-like structure similar in appearance to the southern (S)
component of the VLBI triple at 2.3 GHz (see Fig.~3 of Ulvestad \etal\ 
1999a). Based on its location along the jet axis, we identify this
component as part of the southern jet.  The spectral index between
0.61 and 0.33 GHz of this new southern jet component is steep ($\alpha
= -0.74 \pm 0.2$ where $S_\nu \propto \nu^\alpha$) as would be
expected for such a large (80 mas FWHM) emission region.  This
component is also detected at 1.36 GHz (Fig.~2) as a group of
subcomponents -- the broken-up appearance is probably due to a lack of
sufficiently short baselines at this frequency to adequately sample
this large component.  Furthermore the fact that this component is
most likely unaffected by the free-free absorption suspected towards
the inner disk (see below), indicates that the line-of-sight towards
this component is relatively unobstructed.  One possibility is that
this jet component is on the approaching side in front of the disk,
consistent with the schematic view put forth by Carilli \etal\ (1998).
The smallest scale VLBI structure appears to be a one-sided
sub-relativistic jet pointing to the southwest.  The one-sided
appearance is probably due to free-free absorption on sub-parsec
scales (Ulvestad et al. 1999b); this supports the argument that the
southwestern jet imaged by Ulvestad \etal\ (1999a) is in front of the
disk and leads into the more extended southern VLBI lobe.

\section{Properties of  the Disk in MRK\,231}

\subsection{Diffuse Radio Emission}

Diffuse, extended emission on scales of $\sim$1 kpc is seen in a
number of Seyfert galaxies (Wilson 1988, Ulvestad \& Wilson 1989).  In
fact, based on observations with the partially completed VLA in 1979,
Ulvestad, Wilson \& Sramek (1981) presented evidence of 10--15 mJy of
extended emission around Mrk\,231 at 4.9 GHz, though this emission was
not well localized by those observations.  In Fig.~3, we present
higher resolution observations at 5 -- 22 GHz that clearly show
extended emission in a fairly circular disk centered on the nucleus.
The radial fall-off beyond 200 mas is Gaussian with a FWHM of 860 mas
(980 pc) (Fig.~5).  At all frequencies a north-south extension is
visible in the inner contours.  At lower flux density levels the disk
appears nearly circular, with a possible extension to the East.

\begin{figure}
\vspace{16cm}
\includegraphics{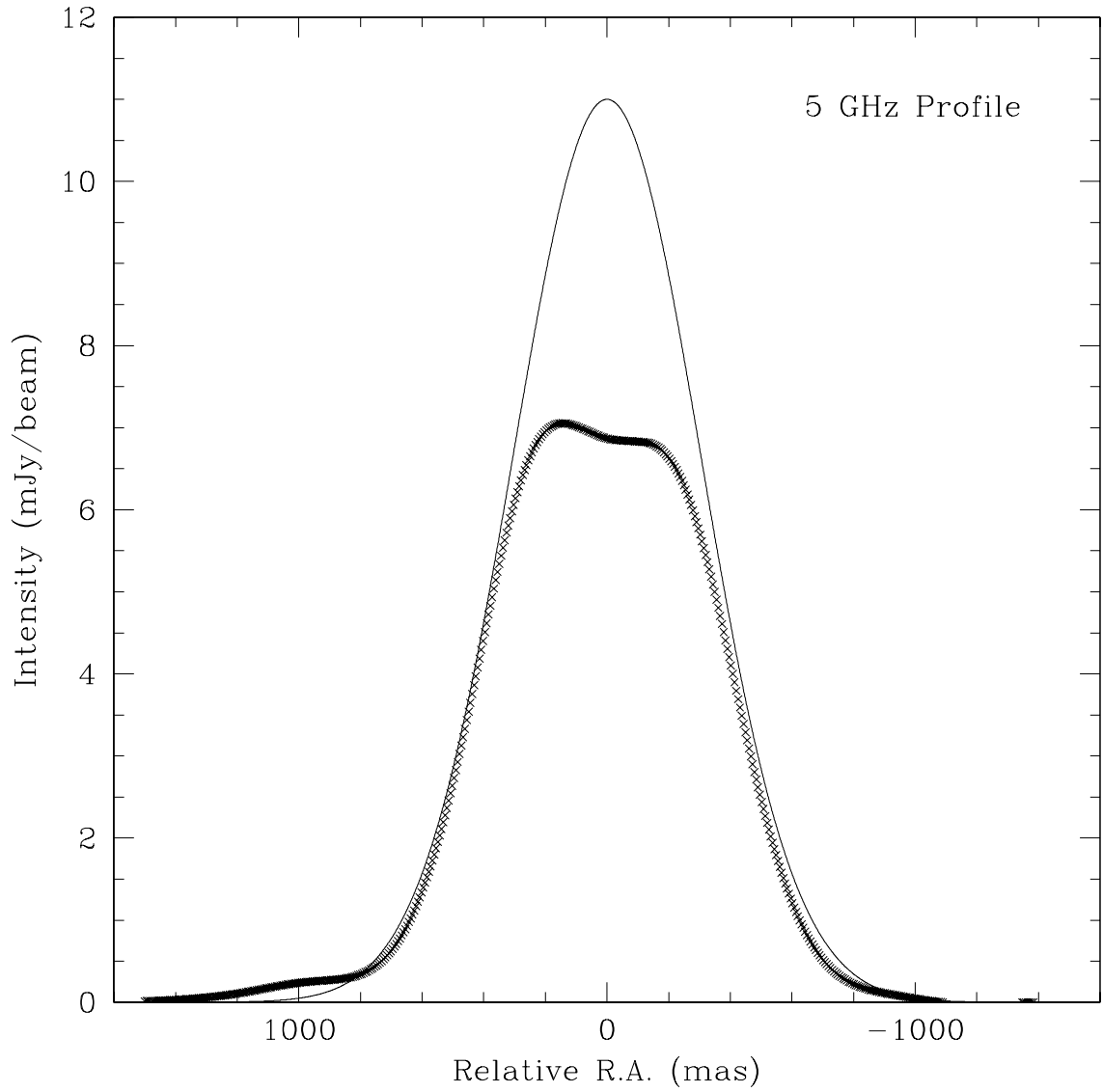}
\figcaption{An east-west slice through the 5 GHz image of the outer 
disk shown in Fig.~3.  The solid line indicates a Gaussian fit with
FWHM 860 mas. 
\label{fig5}}
\end{figure}

\placefigure{fig5}

The existence of a subkiloparsec gas (and stellar?) disk in Mrk\,231 is
now well documented through high resolution molecular line
observations (Bryant \& Scoville 1996, Downes \& Solomon 1998), \HI\ 
21cm absorption line observations, and radio continuum observations
(Carilli \etal\ 1998).  For illustration, we reproduce the results of
the \HI\ 21cm absorption line observations in Fig.~6. This figure shows
the position-velocity diagram along the major axis (oriented
east-west) of the inner disk for \HI\ 21cm absorption. Note the clear
velocity gradient of $\pm$ 130 km s$^{-1}$ to radii $\approx$ 200 mas.
The peak \HI\ optical depth is 0.17. It is likely that the \HI\ extends
farther than is observed in Fig.~6. The spatial limit is set by the
surface brightness of the diffuse radio continuum emission against
which the absorption is seen.  The CO disk seen by Bryant \& Scoville
(1996) on a factor $\sim$3 larger scale has the same velocity gradient
and position angle of the major axis.  Carilli \etal\ (1998) show that
the diffuse radio continuum emission seen on scales from 100 mas to 1
arsecond in Mrk\,231 is likely to be associated with this gas disk, and
in particular with massive star formation. We consider this hypothesis
further below.

\begin{figure}
\vspace{16cm}
\includegraphics{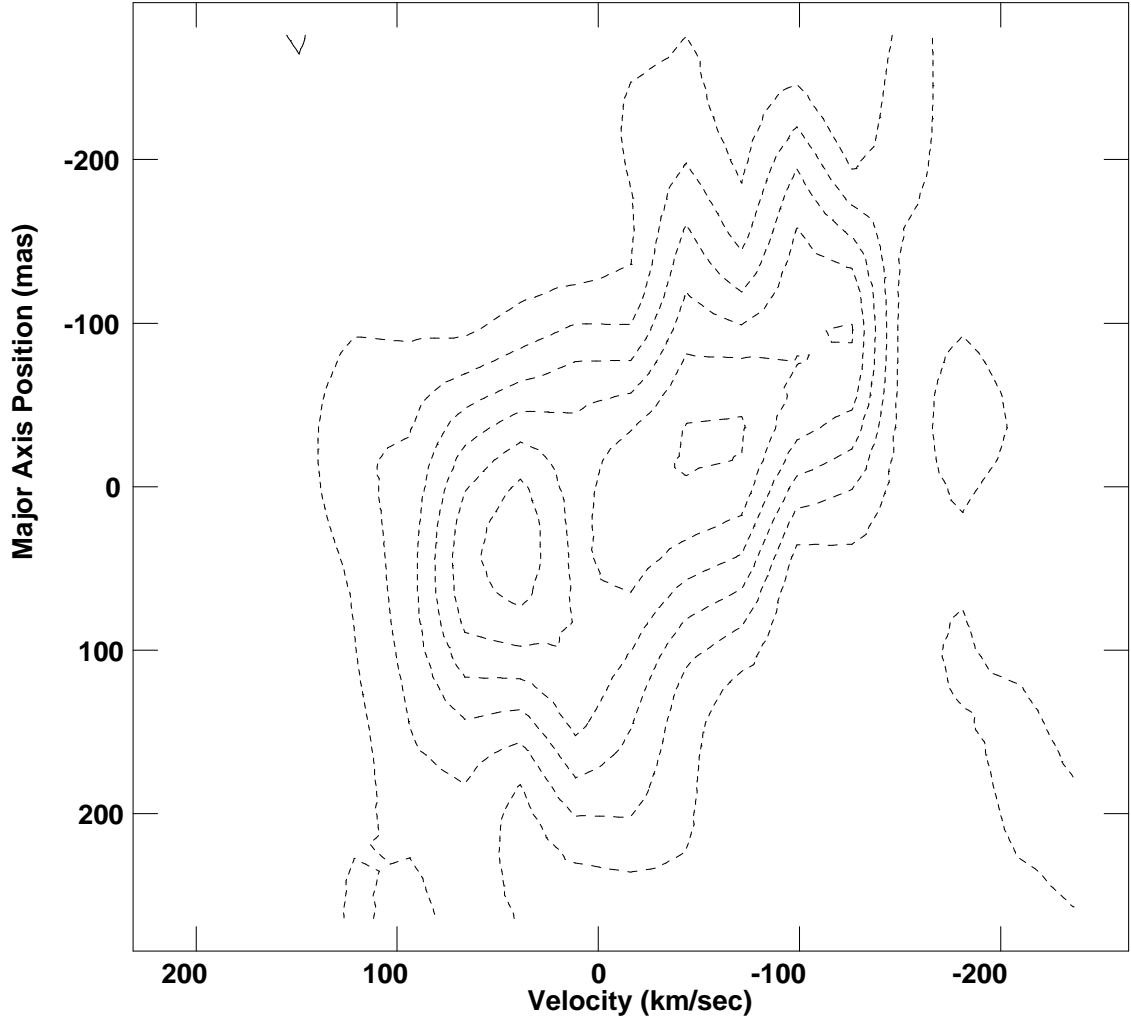}
\figcaption{The \HI\ position-velocity diagram for Mrk\,231 along the major
axis of the disk in position angle 90$^\circ$.
Contours are drawn at $-$6,$-$5,$-$4,$-$3,$-$2,$-$1, and 1 mJy/beam.
The beam has a FWHM of 150 mas. The peak optical depth is 0.17, and
zero velocity corresponds to a heliocentric redshift of 0.04217
(cz = 12642 km s$^{-1}$).  The velocity resolution is 56 km s$^{-1}$.
\label{fig6}}
\end{figure}

\placefigure{fig6}

The radio continuum images presented herein reveal the structure of the
disk at frequencies ranging from 0.33 GHz to 22 GHz. The size of the
disk at 5 GHz (Fig.~3) is $1.9''\times1.6''$ to the 5$\sigma$ surface
brightness level of 0.16 mJy beam$^{-1}$. The  position angle of
the major axis is $\sim$55$^\circ$.
For comparison, the  disk seen in CO
emission has a Gaussian FWHM = $0.9''\times0.8''$ at a position angle of
77$^\circ$. Downes \& Solomon (1998) fit gas kinematic models to the 
CO velocity distribution and find a disk thickness of 23 pc.
Both the radio continuum and CO data imply that the disk must be close
to face-on, with $i \approx 30^\circ$. 

The spectral index of the radio continuum outer disk emission between
5 GHz and 8 GHz is $-$0.7$\pm$0.2 for radii between 0.4$''$ and
0.7$''$. This is consistent with non-thermal synchrotron emission from
a population of relativistic electrons accelerated in shocks driven by
supernova remnants, as would be the case for active star formation in
the disk (Condon 1992).  The IR-to-radio flux density ratio parameter,
Q, for the disk is 2.5 (Carilli \etal\ 1998). This value is consistent
with the value of Q = 2.3$\pm$0.2 seen for the integrated emission
from starforming galaxies (Condon 1992).  The flux density of the disk
at 1.3 GHz is 130 mJy, implying a radio spectral luminosity of
$9\times10^{30}$ ergs s$^{-1}$ Hz$^{-1}$. Using the empirical
relations in Condon (1992) leads to a massive star formation rate in
the disk of 220 M$_\odot$ yr$^{-1}$, and an expected supernova rate of
8 yr$^{-1}$.

An important question is: do the synchrotron emitting electrons
originate in the disk, or are they somehow transported to the disk
from the AGN region?  One key factor to consider is the timescale for
relativistic electron transport compared to the lifetimes of the
relativistic electrons due to synchrotron and
inverse Compton losses. For fields of order 200 $\mu$G (see below),
the expected 
lifetime of the particles radiating at 5 GHz is of order 10$^5$ yrs. 
A standard assumption in ISM plasma physics is that the cosmic ray
electrons are limited to stream at the Alfven velocity due to
scattering off  self-induced Alfven waves -- the streaming energy
is converted into hydromagnetic waves by the two-stream instability
(Wentzel 1974). Using a thermal particle density of 185 cm$^{-3}$ (see
below) implies an Alfven speed of 15 km s$^{-1}$ in the disk, or a
propagation length of only 1.5 pc in 10$^5$ yrs. This short distance
would argue in favor of {\sl in situ} particle acceleration in the
disk. On the other hand, high energy electrons originating in 
solar flares are known to propagate
hyper-Alfvenically in the inter-planetary medium, in violation of the
standard theory (Hudson \& Ryan 1995).  The other extreme is to assume that all
the particles stream at the velocity of light, c, along a tangled
magnetic field. The timescale, $t_o$, for the particles to `random
walk' a distance $R_o$ becomes: $t_o \approx {{R_o^2}\over{l_oc}}$,
where $l_o$ is the cell size for the turbulent field. Using $R_o$
$\approx$ 1$''$ = 1120 pc (= maximum observed radius of the disk at 5
GHz), leads to: $t_o \approx 1\times10^5 [{{l_o}\over{40 \rm
pc}}]^{-1}$ yrs. Hence, if the turbulent magnetic 
field cell size is greater than 40 pc, and if the
particles are somehow allowed to stream at the speed of light, then it
is possible that the diffuse radio continuum emitting regions 
in Mrk\,231 could be populated by electrons from the active nucleus. 

Other arguments in favor of {\sl in situ} particle acceleration in the
disk are: (i) the observed distribution (position angle of major axis,
and major-to-minor axis ratio) of the radio continuum emission is
similar to that seen for the CO disk, (ii) the spectral index of the
extended AGN components between 2.3 GHz and 8.4 GHz is $-$1.5 (Ulvestad
\etal\ 1999a), which is considerably steeper than is seen for the disk,
(iii) the disk emission obeys the IR-radio correlation for star
forming galaxies, and (iv) physical conditions in the disk are
conducive to star formation, and the star formation rate derived from
the molecular line observations is comparable to that derived from the
radio continuum observations (Downes \& Solomon 1998, Bryant \&
Scoville 1996).  Overall, we feel it is most likely that the diffuse
radio continuum emission is driven by star formation in the disk,
although we cannot rule-out an AGN origin for the relativistic
electrons.

\subsection{Free-free absorption of the inner disk}

The integrated spectrum of the inner disk ($\le$ 100 mas) in Mrk\,231
shows an inversion below 1.3 GHz, most likely due to free-free
absorption (Fig.~4). We have fit a free-free absorption model to the
data, and obtain an emission measure, EM = $7.9\pm0.6 \times 10^5 ~
({{\rm T_K}\over{10^4}})^{3\over2}$ pc cm$^{-6}$, where T$_{\rm K}$ is
the kinetic temperature of the gas. Using a disk thickness of 23 pc leads
to n$_e$ = 185 $({{\rm T_K}\over{10^4}})^{3\over4}$ cm$^{-3}$ and
N$_e$ = 1.3$\times10^{22}$ $({{\rm T_K}\over{10^4}})^{3\over4}$
cm$^{-2}$.  This column density is comparable to the hydrogen column
density derived from CO observations (Downes \& Solomon 1998), in \HI\ 
assuming a spin temperature of 1000 K (Carilli \etal\ 1998), and to
the hydrogen column density inferred from soft X-ray absorption of 6
$\times 10^{22}$ cm$^{-2}$ (Nakagawa \etal\ 1997).

The pressure in the ionized absorbing medium is 
5$\times10^{-10}$ $({{\rm T_K}\over{10^4}})^{7\over4}$ dynes
cm$^{-2}$. The pressure in the molecular gas in the disk 
is 5$\times10^{-11}/f$ dynes cm$^{-2}$, where $f$ is the volume
filling factor. Hence, pressures in the ionized and molecular
components are comparable assuming a reasonable filling factor of 0.1
for the molecular gas (Downes \& Solomon 1998). 
The minimum pressure in the relativistic electrons and magnetic fields 
ranges from 2$\times 10^{-10} $ dynes cm$^{-2}$ in the outer disk  
to 10$\times 10^{-10} $ dynes cm$^{-2}$ in the inner disk, making the
standard assumptions of Miley (1980), and in particular using a low
value for the proton-to-electron energy density ratio, ($k \approx$ 1)
and assuming unit filling factor. The corresponding magnetic field
strengths are 80 $\mu$G and 200 $\mu$G, respectively. 

\subsection{Candidate Radio Supernovae}

A final question that can be addressed by these data is the existence
of radio supernovae (RSNe) in the disk of Mrk\,231. Recent VLBI
observations of Arp 220 have revealed 13 RSNe in the inner 100 pc with
flux densities at 1.3 GHz between 0.1 and 1.2 mJy, with three RSNe
between 1.0 mJy and 1.2 mJy (Smith \etal\ 1998). Smith \etal\ show
that these RSNe are of the same class as RSN 1986J observed in the
disk of NGC 891 (Rupen \etal\ 1987): Type II RSNe with luminosities of
order 10$^{28}$ ergs s$^{-1}$ Hz$^{-1}$ and exponential decay times of
3 yrs. The number of RSNe observed in Arp 220 is consistent with a
massive star formation rate of 70 M$_\odot$ yr$^{-1}$.

We have searched for RSNe in Mrk\,231 at a resolution of 10 mas at 1.3
GHz (Fig.~2). The difficulty in Mrk\,231 is confusion of the inner 50
pc due to the AGN radio components.  Based on the observations of Arp
220, and correcting for the relative distances, the predicted number
of RSNe $\ge$ 0.2 mJy in Mrk\,231 is six.  We have identified four
possible RSNe candidates between 0.2 and 0.5 mJy that are not along
the jet or counterjet axis.  The brightest of these four (source A) is
certainly a real source, but it is close to the radio axis and hence
may be related to the AGN. The fainter sources (B,C,D) are farther
away from the radio axis, and are detected at the $\sim$10$\sigma$
level, but we cannot rule out that these `sources' are imaging
artifacts caused by dynamic range limitations imposed by the presence
of the strong AGN components.  The arc of star-forming knots
$\sim$3\arcsec\ south of the core (on the radio-jet axis) seen by
Surace \etal\ (1998) may be jet-induced.  While it is possible that
the jet may trigger star formation along the radio axis closer to the
nucleus, we cannot from our radio observations alone distinguish such
emission from jet emission related to the AGN.  The RSNe in Mrk\,231
hypothesis could be tested by monitoring the system with annual,
sensitive, high quality VLBI imaging to determine the light curves for
the faint possible sources in the star forming disk.

\acknowledgments 
We are grateful to Joan Wrobel, and the referee, J.\ Mazzarella, for
insightful comments.  This research has made use of the NASA/IPAC
Extragalactic Database (NED) which is operated by the Jet Propulsion
Laboratory, Caltech, under contract with NASA.

%
%

\clearpage


\clearpage

\end{document}